\documentclass{PoS}

\usepackage{graphicx}
\usepackage{amsmath}
\usepackage{amssymb}
\usepackage{media9}

\usepackage{tikz,varwidth}
\usetikzlibrary{shapes.geometric}
\usetikzlibrary{arrows.meta,shapes,automata,petri,positioning,calc,decorations.markings}

\title{Visualizations of Centre Vortex Structure in Lattice Simulations}

\ShortTitle{Visualizations of Centre Vortex Structure}
\author{James C. Biddle\\
Centre for the Subatomic Structure of Matter, Department of Physics,\\ The University of Adelaide, SA 5005, Australia\\
        E-mail: \email{james.biddle@adelaide.edu.au}}

\author{Waseem Kamleh\\
Centre for the Subatomic Structure of Matter, Department of Physics,\\ The University of Adelaide, SA 5005, Australia\\
        E-mail: \email{waseem.kamleh@adelaide.edu.au}}

\author{\speaker{Derek B. Leinweber}\thanks{DBL thanks Michael Engelhardt for interesting and
    helpful conversations at Lattice 2018. The authors wish to thank Daniel Trewartha for his
    contributions to the gauge ensembles underlying this investigation.  This research is supported
    with supercomputing resources provided by the Phoenix HPC service at the University of Adelaide
    and the National Computational Infrastructure (NCI) supported by the Australian
    Government. This research is supported by the Australian Research Council through Grants
    No.\ DP140103067, DP150103164, and LE160100051.}\\
Centre for the Subatomic Structure of Matter, Department of Physics,\\ The University of Adelaide, SA 5005, Australia\\
        E-mail: \email{derek.leinweber@adelaide.edu.au}}

\abstract{ 
The structure of centre vortices in $SU(3)$ gauge-field configurations
is examined through modern visualization techniques.
Centre vortices are identified through gauge transformations
maximizing the centre of the gauge group.  Focusing on the thin
vortices identified by Wilson loops having a non-trivial centre phase,
the vortex structure is illustrated through renderings of oriented
spatial plaquettes.  Time oriented plaquettes are illustrated by
identifying spatial links associated with these non-trivial
plaquettes.
The results illustrate an important role for the branching of $SU(3)$
centre vortices, possibly through monopole anti-monopole dynamics.
Of particular interest is the correlation of the vortex structure and
the topological-charge structure of the gauge fields, vital to
dynamical chiral symmetry breaking and its associated mass generation.
Singular points, where the projected vortex structure contains
non-trivial centre phases in associated plaquettes spanning all four
dimensions, are observed within the regions of nontrivial topological
charge density calculated on the original Monte-Carlo generated
configurations.
The results provide new insights into the role of centre vortices in
underpinning non-trivial topology in gauge fields. They reveal how the
removal of centre-vortices necessarily destroys non-trivial topology
and destabilizes would-be instantons under smoothing algorithms.
The observed correlations further strengthen the idea that centre
vortices are the seeds of dynamical chiral symmetry breaking. 
}

\FullConference{The 36th Annual International Symposium on Lattice Field Theory - LATTICE2018\\
		22-28 July, 2018\\
		Michigan State University, East Lansing, Michigan, USA.}

\begin{document}

%% 7 pages total

\section{Introduction}

The essential, fundamentally-important, nonperturbative features of
the QCD vacuum fields are: the dynamical generation of mass through
chiral symmetry breaking, and the confinement of quarks.  But what is
the fundamental mechanism of QCD that underpins these phenomena?  What
aspect of the QCD vacuum causes quarks to be confined?  Which aspect
is responsible for dynamical mass generation?  Do the underlying
mechanisms share a common origin?

Recent research is now exposing the centre-vortex structure of
nonperturbative gluon-field configurations to be the most fundamental
aspect of nonperturbative vacuum structure giving rise to both
confinement and dynamical chiral-symmetry breaking.

Removal of $SU(3)$ centre vortices removes confinement, while
consideration of the vortices alone provides confinement
\cite{Langfeld:2003ev}.  Also, the planar vortex density of
centre-vortex degrees of freedom scales with the lattice spacing
providing an well defined continuum limit \cite{Langfeld:2003ev}.
Similarly, removal of vortices suppresses the infrared enhancement of
the gluon propagator.  Again the vortices alone contain the long
distance structure of the gluon fields responsible for the well-known
infrared enhancement \cite{Biddle:2018dtc}.  Studies of the
nonperturbative quark propagator of the overlap-Dirac fermion operator
further strengthen the fundamental role of centre vortices.  A
connection between center vortices and instantons was established
through gauge-field smoothing \cite{Trewartha:2015ida} and evidence
that centre vortices underpin dynamical chiral symmetry breaking in
$SU(3)$ gauge theory was reported in Ref.~\cite{Trewartha:2015nna}.
Moreover the removal of centre vortex degrees of freedom from the
gluon fields restores chiral symmetry \cite{Trewartha:2017ive}.
Centre vortices are the seeds of dynamical chiral symmetry breaking.

In light of the importance of these most fundamental aspects of QCD
vacuum structure, we will present visualizations of the complex
structures formed by the projected centre vortices in $SU(3)$ gauge
theory and explore their correlation with the topological charge
density of the gluon fields.

\section{Centre Vortex Identification}

Centre vortices are identified through a gauge fixing procedure
designed to bring the lattice link variables as close as possible to
the identity multiplied by phase equal to 
one of the three cube-roots
of one.  

Here, vortices are identified by gauge fixing the original
Monte-Carlo generated configurations directly to Maximal Centre Gauge
\cite{DelDebbio:1996mh,Langfeld:1997jx,Langfeld:2003ev}, without any
preconditioning \cite{Cais:2008za}.
The links $U_\mu(x)$ are gauge transformed to be brought close
to the centre elements of $SU(3)$,
\begin{equation} 
Z = \exp \left ( 2 \pi i\, \frac{m}{3} \right ) \, \mathbf{I}, \textrm{ with } m = -1, 0, 1.
\label{CentreSU3}
\end{equation}
On the lattice this is implemented by searching for the gauge
transformation $\Omega$ such that,
\begin{equation}
\sum_{x,\mu} \,  \left | \mathrm{tr}\, U_\mu^\Omega(x) \, \right |^2 \stackrel{\Omega}{\to}
\mathrm{max} \, .
\label{GaugeTrans}
\end{equation}
One then projects the gluon field to a centre-vortex only
configuration where each link is a number (one of the roots of unity)
times the identity matrix
\begin{equation} 
U_\mu(x) \to Z_\mu(x) \textrm{ where }
Z_\mu(x) = \exp \left ( 2 \pi i\, \frac{m_\mu(x)}{3} \right
)\mathbf{I} \, ,
\label{UtoZ}
\end{equation}
where $m_\mu(x) = -1, 0, 1.$

The vortices are identified by the centre charge, $z$, found
by taking the product of the links around a plaquette,
\begin{equation}
z = \prod_\Box Z_\mu(x) = \exp \left ( 2 \pi i\, \frac{m}{3} \right ) \, .
\label{CentreCharge}
\end{equation}
A right-handed ordering of the dimensions is selected in calculating
the centre charge.  If $z=1$, no vortex pierces the plaquette. If $z
\neq 1$ a vortex with charge $z$ pierces the plaquette.
Vortices can be removed from the original gauge-field configuration by making the transformation
\begin{equation}
U_\mu(x) \to U_\mu^\prime(x) = Z_\mu^*(x)\, U_\mu(x) \, .
\label{RemoveCV}
\end{equation}

\section{Centre Vortex Visualization Methods}

%% A $20^{3} \times 40$ lattice is considered for $SU(3)$ Lu\"scher-Weisz \cite{Luscher:1984xn}
%% mean-field improved gauge fields with $a = 0.125 \, $fm.  Calculations of the topological charge
%% density are performed with an $\mathcal{O}(a^4)$-five-loop improved field-strength tensor,
%% $F_{\mu\nu}$, designed to ensure the most local $1\times 1$ and $1\times 2$ clover terms make up
%% 96\% of the loop contributions \cite{BilsonThompson:2002jk}.  The calculations follow seven sweeps
%% of an $\mathcal{O}(a^4)$-three-loop improved action \cite{BilsonThompson:2003zi} to ensure the
%% lattice operators are accurate while preserving topological structure within the gauge fields.

\subsection{Spatially Oriented Projected Centre Vortices}

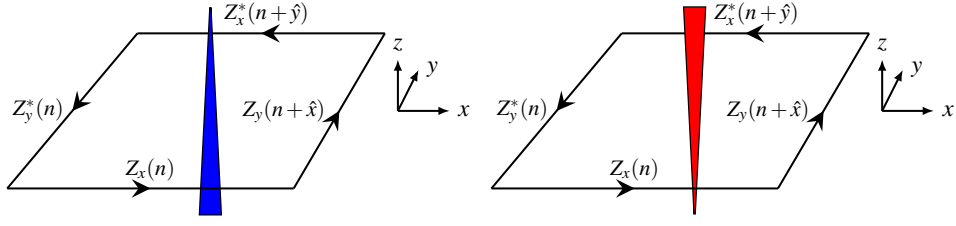
\begin{figure}[t]
\begin{center}
\resizebox{!}{3cm}{%
\begin{tikzpicture}[scale=0.9]
\begin{scope}[very thick,decoration={
    markings,
    mark=at position 0.5 with {\arrow[scale=2]{stealth}}}
    ] 
  % bottom right to top right                    x,y start of line    label  x,y end
  \draw[line width=1.0,postaction={decorate}](1.5,-1.5)-- node[left]{$Z_y(n+\hat x)\ $} (3.25,1.5)node(g){};
  % top right to top left
  \draw[line width=1.0,postaction={decorate}](3.25,1.5)-- node[above]{${}\ \ Z_x^*(n+\hat y)$} (-1.5,1.5);
  % top left to bottom left
  \draw[line width=1.0,postaction={decorate}](-1.5,1.5)-- node[left]{$Z_y^*(n)$}(-4,-1.5);

  % Jet triangle
  % bottom left	
  \draw (-0.3,-2) node(a){}
  -- (0.1,-2) node(b){}   % bottom right
  -- (-0.1,2) node(c){}   % top
  -- cycle;               % complete
  \fill[blue] (a.center) -- (b.center) -- (c.center);
  
  % bottom left to bottom right
  \draw[line width=1.0,postaction={decorate}](-4,-1.5)-- node[above]{$Z_x(n)$}(1.5,-1.5)node(f){};
  
  % Coordinate axes       arrow head          x,y start -- x,y finish [position] label
  \draw[line width=1.0,-{Latex[length=2mm]}](3.5,0)--(4.5,0.0)node[right]{\large $x$};
  \draw[line width=1.0,-{Latex[length=2mm]}](3.5,0)--(3.9,0.8)node[right]{\large $y$};
  \draw[line width=1.0,-{Latex[length=2mm]}](3.5,0)--(3.5,1.0)node[above]{\large $z$};
  \end{scope}
\end{tikzpicture}
}
\resizebox{!}{3cm}{%
\begin{tikzpicture}[scale=0.9]
\begin{scope}[very thick,decoration={
    markings,
    mark=at position 0.5 with {\arrow[scale=2]{stealth}}}
    ] 
  % bottom right to top right                    x,y start of line    label  x,y end
  \draw[line width=1.0,postaction={decorate}](1.5,-1.5)-- node[left]{$Z_y(n+\hat x)\ $} (3.25,1.5)node(g){};
  % top right to top left
  \draw[line width=1.0,postaction={decorate}](3.25,1.5)-- node[above]{\quad $Z_x^*(n+\hat y)$} (-1.5,1.5);
  % top left to bottom left
  \draw[line width=1.0,postaction={decorate}](-1.5,1.5)-- node[left]{$Z_y^*(n)$}(-4,-1.5);

  % Jet triangle
  \draw (-0.3,2) node(a){}
  -- (0.1,2) node(b){}
  -- (-0.1,-2)node(c){}
  -- cycle;
  \fill[red] (a.center) -- (b.center) -- (c.center);
  
  % bottom left to bottom right
  \draw[line width=1.0,postaction={decorate}](-4,-1.5)-- node[above]{$Z_x(n)$}(1.5,-1.5)node(f){};
  
  % Coordinate axes       arrow head          x,y start -- x,y finish [position] label
  \draw[line width=1.0,-{Latex[length=2mm]}](3.5,0)--(4.5,0.0)node[right]{\large $x$};
  \draw[line width=1.0,-{Latex[length=2mm]}](3.5,0)--(3.9,0.8)node[right]{\large $y$};
  \draw[line width=1.0,-{Latex[length=2mm]}](3.5,0)--(3.5,1.0)node[above]{\large $z$};
  \end{scope}
\end{tikzpicture}
}
\end{center}
\vspace{-18pt}
\caption{{\bf Rendering the centre charge} of Eq.~(\protect\ref{CentreCharge}) associated with a
  plaquette in the $x$-$y$ plane at lattice site $n$.  (left) An $m = +1$ vortex with centre charge
  $z = \exp(2\pi i / 3)$ is rendered as a blue jet pointing in the $+\hat z$ direction.  (right) An
  $m = -1$ vortex with centre charge $z = \exp(-2\pi i / 3)$ is rendered as a red jet in the $-\hat
  z$ direction.  }
\label{fig:jets}
\end{figure}

As discussed in the previous section, vortex directions are indicated using a right-handed
coordinate system.  For example, with reference to Eq.~(\ref{CentreCharge}), an $m = +1$ vortex in
the $x$-$y$ plane is plotted in the $+\hat z$ direction as a blue jet.  Similarly, an $m = -1$
vortex in the $x$-$y$ plane is plotted in the $-\hat z$ direction as a red jet.  Figure
\ref{fig:jets} provides an illustration of this assignment.

Because the centre charge transforms to its complex conjugate under permutation of the two
dimensions describing the plaquette, the centre charge can be thought of as a directed flow of
charge $z = \exp(2\pi i / 3)$.  With our assignments above, the propagation of charge $z =
\exp(-2\pi i / 3)$ is in the direction opposite to the rendered jet.  Noting that $\exp(-2\pi i /
3)$ is equivalent to two units of charge $\exp(2\pi i / 3)$, the propagation of two units of centre
charge is also in a direction opposite to the rendered jet.  This ambiguity makes it difficult to
differentiate between vortex branching and monopole contributions \cite{Spengler:2018dxt}.

\subsection{Space-Time Oriented Projected Centre Vortices}

In addition to purely spatial plaquettes, every link in the spatial volume has a forward and
backward time-oriented plaquette associated with it.  Thus, the three jets associated with the
spatial $x$-$y$, $y$-$z$ and $z$-$x$ plaquettes, are complemented by jets in the three forward-time
$x$-$t$, $y$-$t$ and $z$-$t$ plaquettes, and jets in the three backward-time $x$-$t$, $y$-$t$ and
$z$-$t$ plaquettes.

As the previous rendering technique cannot be applied to the hidden time dimension, space-time
oriented P vortices are illustrated by rendering the link in the spatial three-volume associated
with the space-time P vortex.  If a spatial link belongs to a P vortex in a space-time plaquette
then the link is rendered in cyan for an $m = +1$ vortex and orange for an $m = -1$ vortex.  In the
spirit of the right-hand rule applied to the ordering of spatial-plaquette coordinates, the
ordering of space-time coordinates is taken with reference to the four-dimensional Levi-Civita
tensor.  To distinguish whether the space-time plaquette is forward or backward in time, the link
is rendered as a positively-directed arrow for forward space-time plaquettes and as a
negatively-directed arrow for backward space-time plaquettes.

%% As one steps forwards in time, positively-directed links become negatively-directed.

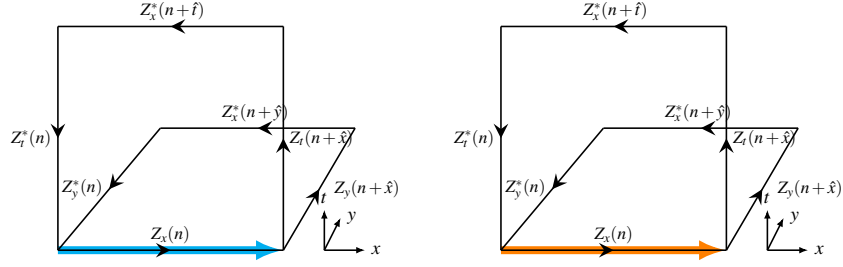
\begin{figure}[t]
\begin{center}
%
%  Colouring a link associated with a time oriented plaquette
%
\resizebox{!}{3.5cm}{%
\begin{tikzpicture}[scale=0.9]
\begin{scope}[very thick,decoration={
    markings,
    mark=at position 0.5 with {\arrow[scale=2]{stealth}}}
    ] 

  % cyan line first to make it look transparent
  \draw[line width=5,color=cyan,-{Latex[length=6mm]}](-4.0,-1.5)--(1.5,-1.5);

  % bottom right to top right                    x,y start of line    label  x,y end
  \draw[line width=1.0,postaction={decorate}](1.5,-1.5)--node[right]{$\ \ Z_y(n+\hat x)\,$} (3.25,1.5)node(g){};
  % top right to top left
  \draw[line width=1.0,postaction={decorate}](3.25,1.5)--node[above]{${}\,\,Z_x^*(n+\hat y)\ \ $} (-1.5,1.5);
  % top left to bottom left
  \draw[line width=1.0,postaction={decorate}](-1.5,1.5)--node[left]{$Z_y^*(n)$} (-4,-1.5);

  %% % Jet triangle
  %% % bottom left	
  %% \draw (-0.3,-2) node(a){}
  %% -- (0.1,-2) node(b){}   % bottom right
  %% -- (-0.1,2) node(c){}   % top
  %% -- cycle;               % complete
  %% \fill[red] (a.center) -- (b.center) -- (c.center);
  
  % bottom left to bottom right
  \draw[line width=1.0,postaction={decorate}](-4.0,-1.5)--node[above]{$Z_x(n)$}(1.5,-1.5)node(f){};
  
  % t direction  
  \draw[line width=1.0,postaction={decorate}]( 1.5,-1.5)--node[right]{$Z_t(n+\hat x)\,$} (1.5,4.0);    % width  3.25-(-1.5) = 4.75
  \draw[line width=1.0,postaction={decorate}]( 1.5, 4.0)--node[above]{${}\,\,Z_x^*(n+\hat t)$}(-4.0,4.0);    % height 6.25-1.5    = 4.75
  \draw[line width=1.0,postaction={decorate}](-4.0, 4.0)--node[left]{$Z_t^*(n)$}(-4.0,-1.5);

  % Coordinate axes       arrow head          x,y start -- x,y finish [position] label
  \draw[line width=1.0,-{Latex[length=2mm]}](2.5,-1.5)--(3.5,-1.5)node[right]{\large $x$};
  \draw[line width=1.0,-{Latex[length=2mm]}](2.5,-1.5)--(2.9,-0.7)node[right]{\large $y$};
  \draw[line width=1.0,-{Latex[length=2mm]}](2.5,-1.5)--(2.5,-0.5)node[above]{\large $t$};
  \end{scope}
\end{tikzpicture}
}
\quad
%
%  Colouring a link associated with a time oriented plaquette
%
\resizebox{!}{3.5cm}{%
\begin{tikzpicture}[scale=0.9]
\begin{scope}[very thick,decoration={
    markings,
    mark=at position 0.5 with {\arrow[scale=2]{stealth}}}
    ] 

  % orange line first to make it look transparent
  \draw[line width=5,color=orange,-{Latex[length=6mm]}](-4.0,-1.5)--(1.5,-1.5);

  % bottom right to top right                    x,y start of line    label  x,y end
  \draw[line width=1.0,postaction={decorate}](1.5,-1.5)--node[right]{$\ \ Z_y(n+\hat x)\,$} (3.25,1.5)node(g){};
  % top right to top left
  \draw[line width=1.0,postaction={decorate}](3.25,1.5)--node[above]{${}\,\,Z_x^*(n+\hat y)\ \ $} (-1.5,1.5);
  % top left to bottom left
  \draw[line width=1.0,postaction={decorate}](-1.5,1.5)--node[left]{$Z_y^*(n)$} (-4,-1.5);

  %% % Jet triangle
  %% % bottom left	
  %% \draw (-0.3,-2) node(a){}
  %% -- (0.1,-2) node(b){}   % bottom right
  %% -- (-0.1,2) node(c){}   % top
  %% -- cycle;               % complete
  %% \fill[red] (a.center) -- (b.center) -- (c.center);
  
  % bottom left to bottom right
  \draw[line width=1.0,postaction={decorate}](-4.0,-1.5)--node[above]{$Z_x(n)$}(1.5,-1.5)node(f){};
  
  % t direction  
  \draw[line width=1.0,postaction={decorate}]( 1.5,-1.5)--node[right]{$Z_t(n+\hat x)\,$} (1.5,4.0);    % width  3.25-(-1.5) = 4.75
  \draw[line width=1.0,postaction={decorate}]( 1.5, 4.0)--node[above]{${}\,\,Z_x^*(n+\hat t)$}(-4.0,4.0);    % height 6.25-1.5    = 4.75
  \draw[line width=1.0,postaction={decorate}](-4.0, 4.0)--node[left]{$Z_t^*(n)$}(-4.0,-1.5);

  % Coordinate axes       arrow head          x,y start -- x,y finish [position] label
  \draw[line width=1.0,-{Latex[length=2mm]}](2.5,-1.5)--(3.5,-1.5)node[right]{\large $x$};
  \draw[line width=1.0,-{Latex[length=2mm]}](2.5,-1.5)--(2.9,-0.7)node[right]{\large $y$};
  \draw[line width=1.0,-{Latex[length=2mm]}](2.5,-1.5)--(2.5,-0.5)node[above]{\large $t$};
  \end{scope}
\end{tikzpicture}
}
\end{center}
\vspace{-12pt}
\caption{{\bf Rendering the centre charge} of Eq.~(\protect\ref{CentreCharge}) associated with a
  space-time plaquette at lattice site $n$.  An $m = +1$ vortex with centre charge $z = \exp(2\pi i
  / 3)$ is rendered as a cyan link in the spatial volume (left) whereas an $m = -1$ vortex with
  centre charge $z = \exp(-2\pi i / 3)$ is rendered as an orange link in the spatial volume
  (right).  The directions of the arrow rendered indicate whether the space-time plaquette is
  forward or backward in time. Here the arrows point in the $+\hat x$ direction as a
  forward-looking plaquette is under consideration.  }
\label{fig:lines}
\end{figure}

\section{Centre Vortex Structure}

Projected centre vortices (P-vortices) are surfaces in four dimensional space-time, analogous to
the centre line of a vortex in fluid dynamics that maps out a surface as it moves through time.  As
the surface cuts through the three-dimensional volume of our visualization, a P-vortex line mapping
the flow of centre charge is rendered.

\begin{figure}[t]
\begin{center}
\includemedia[
        noplaybutton,
	3Dtoolbar,
	3Dmenu,
	3Dviews=Plaq_CFG95_T02.vws,
	3Dcoo  = 10 10 20, %Centre of orbit, half lattice length in each direction
	3Dc2c  = 0 1 0,    % Direction vector from camera to centre of orbit
	3Droo  = 70,       % Radius of orbit, distance from camera to centre of orbit
	3Droll = 270,      % Roll about the camera vector in degrees
	3Dlights=CAD,
	width=\textwidth,      % for 1330x970 from full screen capture
]{\includegraphics{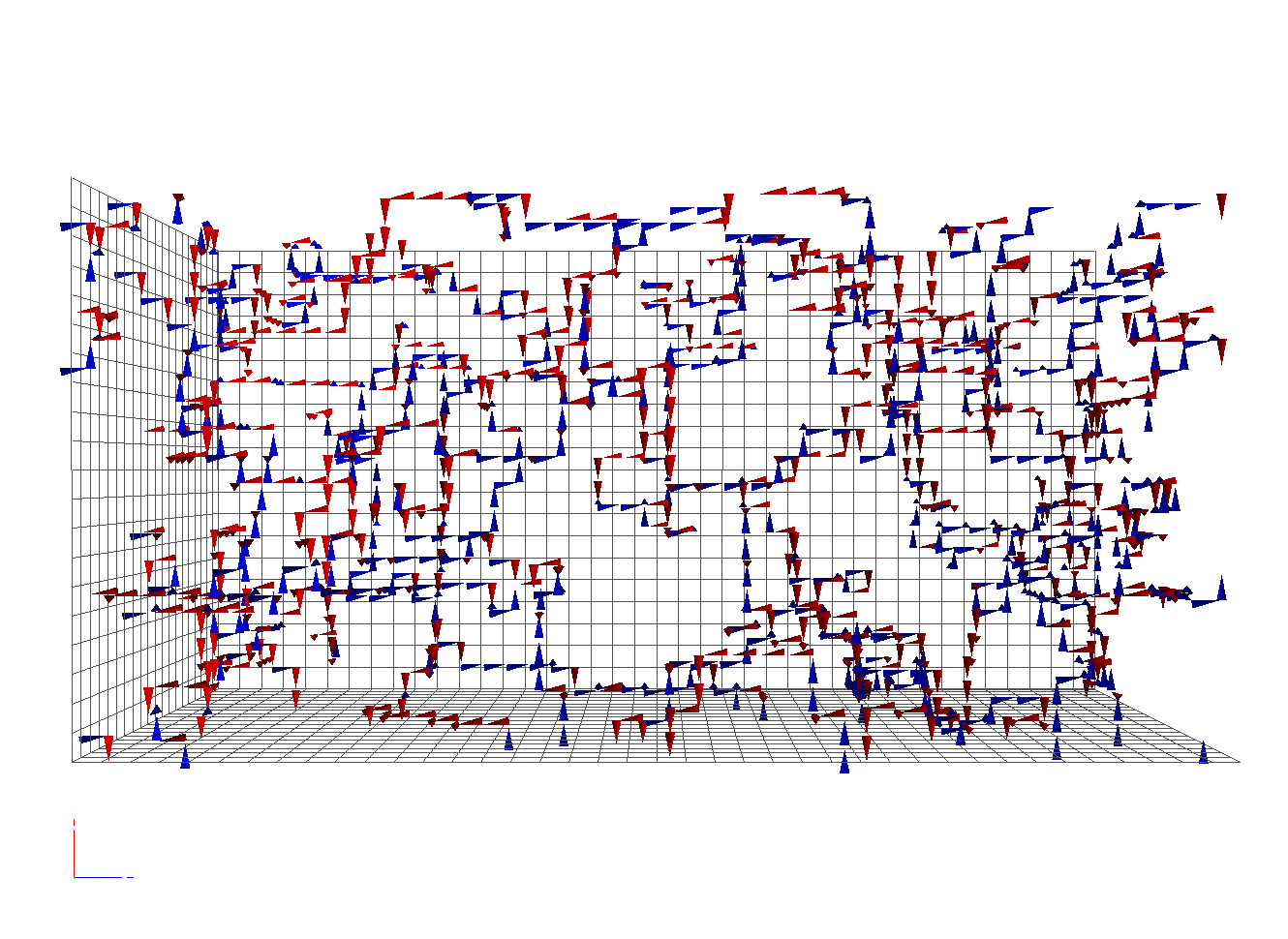}}{Plaq_CFG95_T02.u3d}
\end{center}
\vspace{-24pt}
\caption{ {\bf The projected centre-vortex structure of the gluon field.}  The flow of centre
  charge is illustrated by the jets as described in the text.  (Click on the image to activate it
  in Adobe Reader. Click and drag to rotate, Ctrl-click to translate, Shift-click or mouse wheel to
  zoom. Right click to access the ``Views'' menu.)  }
\label{Plaq_CFG95_T02.u3d}
\end{figure}

\begin{figure}[t]
\begin{center}
\includemedia[
        noplaybutton,
	3Dtoolbar,
	3Dmenu,
	3Dviews=PlaqLink_CFG95_T02.vws,
	3Dcoo  = 10 10 20, %Centre of orbit, half lattice length in each direction
	3Dc2c  = 0 1 0,    % Direction vector from camera to centre of orbit
	3Droo  = 70,       % Radius of orbit, distance from camera to centre of orbit
	3Droll = 270,      % Roll about the camera vector in degrees
	3Dlights=CAD,
	width=\textwidth, % for 1330x970
]{\includegraphics{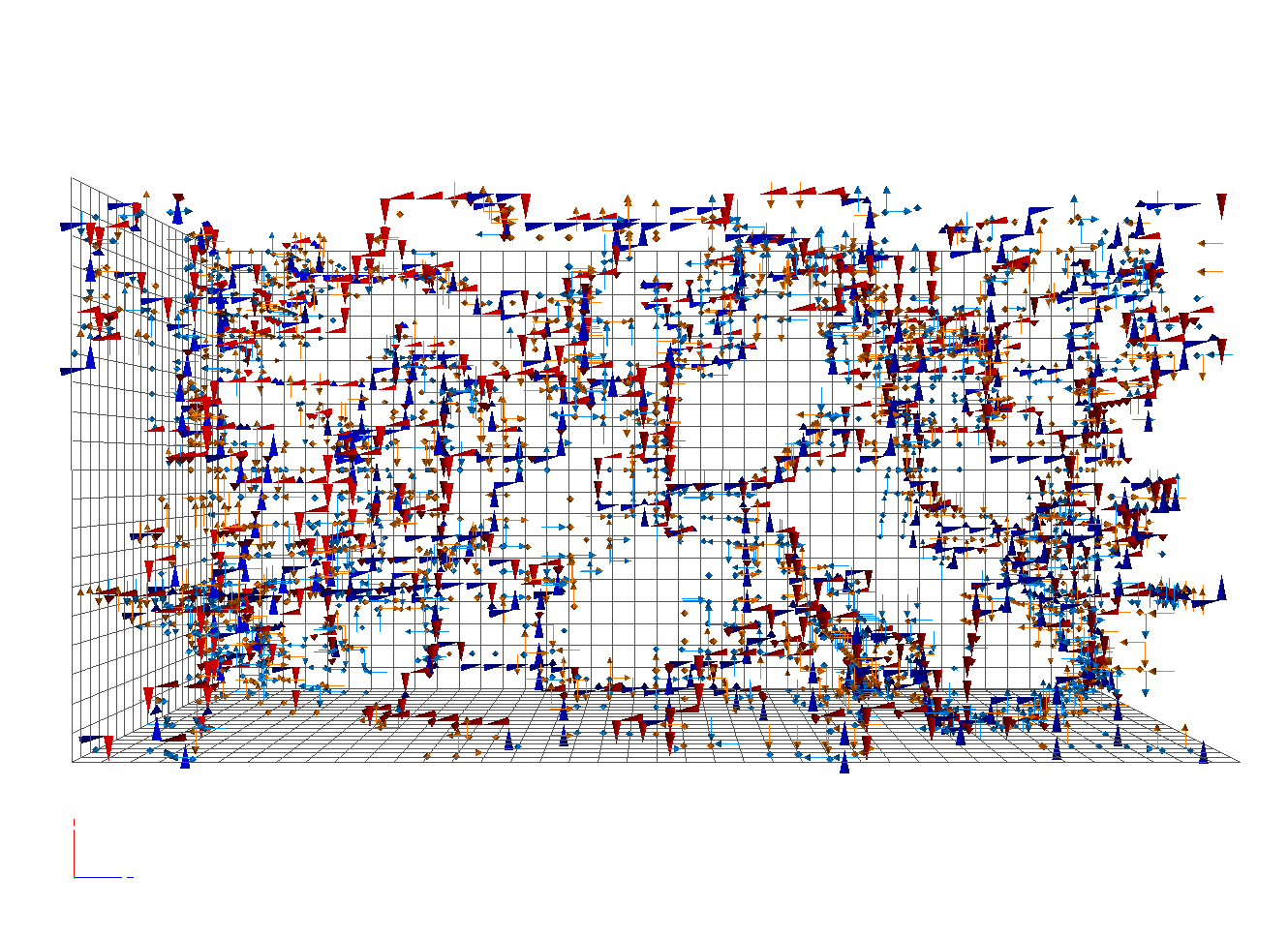}}{PlaqLink_CFG95_T02.u3d}
\end{center}
\vspace{-24pt}
\caption{{\bf The centre-vortex structure of a gluon field configuration from both spatial and
  space-time oriented plaquettes.}  The flow of the centre charge of spatial plaquettes is illustrated
  by the jets and the spatial links of space-time oriented plaquettes, indicating the motion of the
  jets through time, are rendered in orange and cyan as described in the text.
  (Click to activate the image in Adobe Reader.  Controls are summarised in the caption of
  Fig.~\protect\ref{Plaq_CFG95_T02.u3d}.)}
\label{PlaqLink_CFG95_T02.u3d}
\end{figure}

The structure of projected centre vortices associated with spatial plaquettes is
illustrated\footnote{To interact with Figs.~\ref{Plaq_CFG95_T02.u3d} and \ref{PlaqLink_CFG95_T02.u3d},
%  and \ref{PlaqLinkTopQ_CFG95_T02.u3d}, 
  open this pdf document in Adobe Reader 9 or later.  Linux
  users should install \href{ftp://ftp.adobe.com/pub/adobe/reader/unix/9.x/9.4.1/enu/}{Adobe
    acroread version 9.4.1}, the last edition to have full 3D support.  From the ``Edit'' menu,
  select ``Preferences...'' and ensure ``3D \& Multimedia'' is enabled and ``Enable double-sided
  rendering'' is selected.} in Fig.~\ref{Plaq_CFG95_T02.u3d}.  
Here, a $20^{3} \times 40$ lattice is considered for $SU(3)$ Lu\"scher-Weisz \cite{Luscher:1984xn}
mean-field improved gauge fields with $a = 0.125 \, $fm.
Inspection of the vortices reveals the flow of centre charge, intersection points and a prevalence
of branching points resembling monopole or anti-monopole contributions, where three jets emerge
from or converge to a point.

Upon activating the image of Fig.~\ref{Plaq_CFG95_T02.u3d} views of these P-vortex features can
be accessed from the Right-click ``Views'' menu including an ``Intersection Point,'' a
``Monopole/Branch Example,'' an ``Anti-Monopole/Branch Example'' and a ``Cluster of Vortex
Branches.''

Figure~\ref{Plaq_CFG95_T02.u3d} displays only one-third of the vortex information associated with
these spatial links.  Figure~\ref{PlaqLink_CFG95_T02.u3d} introduces the P-vortex information
associated with space-time plaquettes.  This new information indicates the motion of vortex lines
from one spatial slice in time to adjacent spatial slices in time, with forward (backward) pointing
arrows indicating the direction of the vortex sheet as one moves forward (backward) in time.  

It is possible to find sheets of links rendered as the vortex sheet runs approximately parallel to
the time dimension.  In this case the movement of the spatial jets can be over several lattice
spacings as one moves from one time slice to the next.  An example of this sheet of links is
available in the right-click ``Views'' menu as ``Vortex Sheet Between Spatial Planes.''

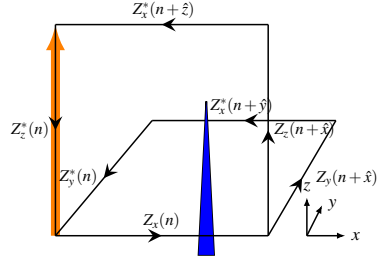
\begin{figure}[t]
\begin{center}
%
%  Singular point signature  (any colour, any corner any direction)
%
\resizebox{!}{3.5cm}{%
\qquad
\begin{tikzpicture}[scale=0.9]
\begin{scope}[very thick,decoration={
    markings,
    mark=at position 0.5 with {\arrow[scale=2]{stealth}}}
    ] 

  % orange line first to make it look transparent
  \draw[line width=6,color=orange,-{Latex[length=6mm]}](-4.0,-1.5)--(-4.0,4.0);

  % bottom right to top right                    x,y start of line    label  x,y end
  \draw[line width=1.0,postaction={decorate}](1.5,-1.5)--node[right]{$\ \ Z_y(n+\hat x)\,$} (3.25,1.5)node(g){};
  % top right to top left
  \draw[line width=1.0,postaction={decorate}](3.25,1.5)--node[above]{${}\,\,Z_x^*(n+\hat y)\ \ $} (-1.5,1.5);
  % top left to bottom left
  \draw[line width=1.0,postaction={decorate}](-1.5,1.5)--node[left]{$Z_y^*(n)$} (-4,-1.5);

  % Jet triangle
  % bottom left	
  \draw (-0.3,-2) node(a){}
  -- (0.1,-2) node(b){}   % bottom right
  -- (-0.1,2) node(c){}   % top
  -- cycle;               % complete
  \fill[blue] (a.center) -- (b.center) -- (c.center);
  
  % bottom left to bottom right
  \draw[line width=1.0,postaction={decorate}](-4,-1.5)--node[above]{$Z_x(n)$}(1.5,-1.5)node(f){};
  
  % z direction  
  \draw[line width=1.0,postaction={decorate}]( 1.5,-1.5)--node[right]{$Z_z(n+\hat x)\,$} (1.5,4.0);    % width  3.25-(-1.5) = 4.75
  \draw[line width=1.0,postaction={decorate}]( 1.5, 4.0)--node[above]{${}\,\,Z_x^*(n+\hat z)$}(-4.0,4.0);    % height 6.25-1.5    = 4.75
  \draw[line width=1.0,postaction={decorate}](-4.0, 4.0)--node[left]{$Z_z^*(n)$}(-4.0,-1.5);

  % Coordinate axes       arrow head          x,y start -- x,y finish [position] label
  \draw[line width=1.0,-{Latex[length=2mm]}](2.5,-1.5)--(3.5,-1.5)node[right]{\large $x$};
  \draw[line width=1.0,-{Latex[length=2mm]}](2.5,-1.5)--(2.9,-0.7)node[right]{\large $y$};
  \draw[line width=1.0,-{Latex[length=2mm]}](2.5,-1.5)--(2.5,-0.5)node[above]{\large $z$};
  \end{scope}
\end{tikzpicture}
}
\end{center}
\vspace{-15pt}
\caption{{\bf Signature of a singular point.}  Singular points are points at which the tangent
  vectors of the vortex surface(s) span all four dimensions.  Thus, singular points are lattice
  sites associated with both a rendered spatial-plaquette jet and a rendered space-time plaquette
  link where the link is parallel to the jet.  In this case, the blue jet is associated with centre
  charge in the $x$-$y$ dimensions and the orange link is associated with centre charge in the
  $z$-$t$ dimensions.  Thus the vortex surface spans all four dimensions at site $n$.
\vspace{-3pt}
}
\label{fig:singular}
\end{figure}

Another point of interest is the presence of singular points \cite{Engelhardt:2000wc} which can generate
nontrivial topological charge density
\begin{equation}
q(x) = \frac{g^2}{32 \pi^2 }\, \epsilon_{\mu \nu \rho \sigma}
 \, F_{\mu \nu}^{ab}(x) \,
 F_{\rho \sigma}^{ba} (x) \, .
\label{TopQ}
\end{equation}
The presence of $\epsilon_{\mu \nu \rho \sigma}$ indicates that the vortex surface must span all
four dimensions at a particular lattice point.  Singular points are points at which the tangent
vectors of the vortex surface(s) span all four dimensions and can include touching points,
intersection points and writhing points \cite{Engelhardt:2000wc,Engelhardt:2010ft}.  In the visualization of
Fig.~\ref{PlaqLink_CFG95_T02.u3d}, singular points are lattice sites associated with both a rendered
spatial-plaquette jet and a rendered space-time plaquette link where the link is parallel to the
jet.  Figure~\ref{fig:singular} presents an example of a singular point.  In short, one searches
for a spatial plaquette with a jet, and a link on the corner parallel to the jet.  The right-click
``Views'' menu of Fig.~\ref{PlaqLink_CFG95_T02.u3d} contains examples of a ``Singular Point'' and a
``Collection of Singular Points.''

Future work will present a correlation between these singular points and the topological charge
density observed in the original configurations.
% \cite{BilsonThompson:2002jk,BilsonThompson:2003zi}.
It is common to see large complex vortex structures at the heart of regions having significant
topoological charge density.  Thus one now has a microscopic understanding of how the removal of
centre-vortices 
%% from the original gauge fields 
via Eq.~(\ref{RemoveCV}) necessarily destroys
non-trivial topology and destabilizes would-be instantons under smoothing algorithms.

\section{Outlook}

The projected centre-vortex structure of $SU(3)$ gauge fields is complicated.  The qualitative idea
of a two-dimensional vortex sheet in four dimensions passing through a three-dimensional
visualisation is realised.  Perhaps the key new insight is the abundance of vortex
branching/monopole-antimonopole dynamics.

%% We have also observed that projected vortices spanning all four dimensions are associated with
%% the positions of topological charge.  It is now clear how the removal of centre vortices
%% necessarily spoils instanton-like structure.

Future directions of study include visualisations of Gribov-copy issues in vortex identification
and explorations of the approach to the continuum limit.  We are also interested in understand the
impact of dynamical fermions \cite{Kamleh:2007ud} on centre-vortex structure.

%\bibliography{VortexStructure}

\providecommand{\href}[2]{#2}\begingroup\raggedright\endgroup

\end{document}